\documentclass[runningheads,a4paper]{llncs}
\usepackage[T1]{fontenc}
\usepackage{graphicx}
\usepackage{amsmath}
\usepackage{comment}
\usepackage[colorlinks=true,urlcolor=blue,citecolor=.,linkcolor=.]{hyperref}
\usepackage{csquotes}
\usepackage{listings}
\usepackage[noabbrev]{cleveref}
\usepackage{microtype}
\usepackage{hyphenat}
\usepackage{tabulary}
\usepackage{array}
\usepackage{xspace}
\usepackage{makecell}
\usepackage[american]{babel}
\usepackage{orcidlink}
\usepackage{tikz}
\newcommand*\circled[1]{\tikz[baseline=(char.base)]{
            \node[shape=circle,draw,inner sep=1pt] (char) {#1};}}
\usepackage{enumitem}
\newlist{rqs}{enumerate}{2}
\setlist[rqs,1]{label=RQ\arabic*.,ref=RQ\arabic*}
\setlist[rqs,2]{label=(\alph*),ref=\thequestionsi(\alph*)}
\usepackage[acronym]{glossaries-extra}
\newabbreviation{rag}{RAG}{Retrieval Augmented Generation}
\newabbreviation{llm}{LLM}{Large Language Model}
\newabbreviation{soc}{SOC}{Service-Oriented Computing}
\newabbreviation{uddi}{UDDI}{Universal Description, Discovery, and Integration}
\newabbreviation{ubr}{UBR}{UDDI Business Registry}
\newabbreviation{nlp}{NLP}{Natural Language Processing}
\newabbreviation{is}{IS}{Information System}
\newabbreviation{ise}{ISE}{Information System Engineering}
\def \theosmodelsmall {\texttt{bge-small-en-v1.5}\xspace}
\def \theosmodellarge {\texttt{NV-Embed-v1}\xspace}
\def \theagent {Discovery Agent\xspace}
\def \therag {OpenAPI RAG\xspace}
\begin{document}
\title{Advanced System Integration: Analyzing OpenAPI Chunking for Retrieval-Augmented Generation}
\titlerunning{Analyzing OpenAPI Chunking for Retrieval-Augmented Generation}
\author{Robin D. Pesl\inst{1}\,\orcidlink{0000-0002-5980-9395} \and
Jerin G. Mathew\inst{2}\,\orcidlink{0000-0002-4626-826X} \and
Massimo Mecella\inst{2}\,\orcidlink{0000-0002-9730-8882} \and
Marco Aiello\inst{1}\,\orcidlink{0000-0002-0764-2124}}
\institute{University of Stuttgart, Stuttgart, Germany\\
\email{\{robin.pesl,marco.aiello\}@iaas.uni-stuttgart.de} \and
Sapienza Università di Roma, Rome, Italy\\
\email{\{mathew,mecella\}@diag.uniroma1.it}
}
\authorrunning{Pesl et al.}
\maketitle              
\begin{abstract}
Integrating multiple (sub-)systems is essential to create advanced \glsxtrfullpl{is}.
Difficulties mainly arise when integrating dynamic environments across the \glsxtrshort{is} lifecycle, e.g., services not yet existent at design time.
A traditional approach is a registry that provides the API documentation of the systems' endpoints.
\glsxtrfullpl{llm} have shown to be capable of automatically creating system integrations (e.g., as service composition) based on this documentation but require concise input due to input token limitations, especially regarding comprehensive API descriptions.
Currently, it is unknown how best to preprocess these API descriptions.
Within this work, we (i) analyze the usage of \glsxtrfull{rag} for endpoint discovery and the chunking, i.e., preprocessing, of state-of-practice OpenAPIs to reduce the input token length while preserving the most relevant information.
To further reduce the input token length for the composition prompt and improve endpoint retrieval, we propose (ii) a \theagent that only receives a summary of the most relevant endpoints and retrieves specification details on demand.
We evaluate \glsxtrshort{rag} for endpoint discovery using the RestBench benchmark, first, for the different chunking possibilities and parameters measuring the endpoint retrieval recall, precision, and F1 score.
Then, we assess the \theagent using the same test set.
With our prototype, we demonstrate how to successfully employ \glsxtrshort{rag} for endpoint discovery to reduce the token count.
While revealing high values for recall, precision, and F1, further research is necessary to retrieve all requisite endpoints.
Our experiments show that for preprocessing, \glsxtrshort{llm}-based and format-specific approaches outperform na\"ive chunking methods.
Relying on an agent further enhances these results as the agent splits the tasks into multiple fine granular subtasks, improving the overall \glsxtrshort{rag} performance in the token count, precision, and F1 score.

\keywords{Retrieval augmented generation \and Large language models \and OpenAPI \and Endpoint discovery \and RestBench.}
\end{abstract}
\section{Introduction}
\label{sec:intro}

OpenAPI is the state-of-practice for describing interfaces for integrating \glspl{is}.
It contains formal elements like paths and natural language constituents like descriptions.
For integrating these systems automatically, automated service composition using \glspl{llm} has been proposed~\cite{pesl2024verfahren,pesl2024uncovering,pesl2024compositio}.
These approaches exploit the capabilities of \glspl{llm} to process formal and natural language input, combining them with the inherent nature of automated service composition of decoupling and independent lifecycle management.
While prohibiting any manual modeling effort by relying on already broadly available OpenAPIs, the approaches face the challenge of limited input token length~\cite{pesl2024compositio}.
This bounds the quantity and extent of the input service description.
Even for proprietary models with a large input token context, e.g., OpenAI's GPT4 with a context size of 128,000 tokens~\cite{openai2024contextsize}, an economic constraint emerges as these models are paid by input and output token count.
Therefore, a smaller prompt length is beneficial to (i) insert further service documentation and (ii) reduce costs for proprietary models.

To address these challenges, \gls{rag}~\cite{lewis2020retrieval} has emerged as a promising resort.
In this approach, the external information is collected in a database, typically structured as a set of documents or document chunks.
The primary goal is retrieving only a small subset of the most relevant documents or document chunks, which is then inserted into the prompt~\cite{lewis2020retrieval}.
How to optimally apply \gls{rag} for endpoint discovery in \gls{is} is open to investigation, leading to the following research questions:
\begin{rqs}[topsep=0pt,align=left,wide=0pt,leftmargin=*]
    \item How best to preprocess, i.e., chunk, OpenAPIs for \gls{rag} endpoint discovery?
    \item Can \gls{llm} agents be employed to reduce token count further and improve retrieval performance?
\end{rqs}

To answer RQ1, we develop an \textit{\therag} system that takes as input service descriptions.
We apply different token-based and \gls{llm}-based chunking strategies to split the documentation and evaluate them based on retrieval quality.
The token-based strategies process the document using a classical parser and then split the parts, e.g., endpoints, into equal-sized chunks.
The \gls{llm}-based strategies let an \gls{llm} create a description, i.e., a summary or a question, for each endpoint and then use these descriptions for similarity matching.
We employ mainstream open-source and proprietary embedding models for similarity matching, which can create an embedding vector for an input.
The similarity between two inputs can then be determined by comparing their embedding vectors using, e.g., the cosine similarity.
We evaluate the \therag and the different chunking strategies by relying on the already available RestBench benchmark~\cite{song2023restgpt} for \glspl{llm} agents, measuring recall, precision, and F1 score for each chunking strategy.
The benchmark consists of the OpenAPI descriptions of Spotify and TMDB and corresponding queries, each with a set of endpoints as the sample solution.

To address RQ2, we propose an \gls{llm} agent called \textit{\theagent}.
As \gls{llm} agents allow the usage of external tools, we first investigate using one tool that simply inputs the results of the \gls{rag} to the prompt.
Then, we experiment with using two tools: the first tool filters and enters the \gls{llm} endpoint summaries to the prompt using \gls{rag}, while the second tool allows the retrieval of the endpoint details on demand.
We resort to the same RestBench benchmark for evaluation and measure recall, precision, F1 score, and additional token count.
As the chunking strategy, we rely on the \gls{llm}-based summary strategy with OpenAI's \texttt{text-embedding-3-large} embedding model~\cite{openai2024embeddings}.

The remainder of the paper is structured as follows.
First, we provide an overview of related works regarding service discovery and \glspl{llm} in \Cref{sec:related_work}.
Then, we present how to use \gls{rag} for endpoint discovery and the OpenAPI chunking strategies in \Cref{sec:design}.
We evaluate and discuss the \gls{rag} and the different chunking strategies in \Cref{sec:evaluation}.
Finally, we conclude with \Cref{sec:conclusion}.

\section{Related Work}
\label{sec:related_work}

Regarding endpoint discovery, we provide a brief overview of the essential concepts of the various service discovery approaches.
Additionally, we provide relevant insights into \glspl{llm} and the novel approach of integrating \glspl{llm} with tools, known as \gls{llm} agents, and how they relate to our approach.

\subsection{Service Discovery}

The most common service discovery implementation is a service registry, which collects information about available services and offers search facilities.
This service registry is usually backed by a component residing at the middleware or application level~\cite{lemos2015web}.
It is characterized by the syntax used to describe the services and their invocation and the expressive power of the available query language.
The typical integration model is a pull model where service consumers search for the required services.
Less is a push model as used in the UPnP protocol, where service providers regularly advertise their services~\cite{santana2006upnp}.

In the early days of XML-based services, the infrastructure for service discovery was the \gls{uddi} specification~\cite{curbera2002unraveling}.
\gls{uddi} had a global incarnation called the \gls{ubr}, intended to offer an Internet-wide repository of available web services and promoted by IBM, Microsoft, and SAP.
Unfortunately, \gls{ubr} never gained widespread adoption and was short-lived (2000-2006).
Significant research in the early days focused on enhancing service discovery on \gls{uddi}, improving search capabilities, and creating federated registries, e.g.,~\cite{baresi2006distributed,bohn2008dynamic,10.1007/978-3-540-24593-3_5}.
Alternatively, WS-Discovery is a multicast protocol that finds web services on a local network.

Nowadays, OpenAPI is the de facto standard for describing services.
While not offering a discovery protocol and mechanism, given its popularity, OpenAPI would also benefit from discovery~\cite{10.1007/978-3-031-57853-3_3}.
So, additional infrastructure for discovery has been proposed, such as centralized repositories (SwaggerHub or Apiary), service registry integration (Consul, Eureka), API Gateways (Kong, Apigee), or Kubernetes annotations (Ambassador).

Populating registries of services requires effort from service providers, which often hinders the success of such approaches, especially if the service provider is expected to provide extensive additional information beyond the service endpoints.
This additional effort has often been the reason for the failure of some of these technologies, most notably \gls{ubr}.
Approaches confined to specific applications, domains, or enterprises have been more successful, e.g., Eureka.
Developed by Netflix as part of its microservices architecture~\cite{thones2015microservices}, Eureka helps clients find service instances described by host IP, port, health indicator URL, and home page.
Developers can add additional data to the registry for extra use cases.

While classical incarnations like \gls{uddi} used to be comprehensive, they required extensive modeling, e.g., as semantic annotations.
Hence, our approach relies on already broadly available state-of-practice OpenAPI specification and their integration with \gls{rag}.

\subsection{\glsxtrlongpl{llm}}
\glspl{llm} represent one of the recent advancement in the \gls{nlp} and machine learning field~\cite{achiam2023gpt,llama3modelcard,kim2024leveraging}.
Often containing billions of parameters, these models are trained on extensive text corpora to generate and manipulate human-like text~\cite{radford2019better}.
They are primarily based on an encoder-decoder architecture called Transformers~\cite{vaswani2017attention}, which has been further refined to improve text generation tasks using decoder-only models such as GPT~\cite{radford2018improving}.
Usually, the input is a natural language task called prompt, which first needs to be translated to a sequence of input tokens.
The model processes this prompt and returns an output token sequence, which can then be translated back to a natural language answer.
As these models can, in general, capture intricate linguistic nuances and semantic contexts, they can be applied to a wide range of tasks, e.g., in software engineering~\cite{fan2023large}.
\glspl{llm} can be used to create integration based on endpoint documentation automatically~\cite{pesl2024verfahren,pesl2024uncovering,pesl2024compositio}.
Yet, these face strict input token limitations, e.g., 128,000 tokens for current OpenAI models~\cite{openai2024contextsize,pesl2024compositio}.
With this paper, we analyze how \gls{rag} can be used to preprocess API documentation to mitigate this issue.

Another approach is encoder-only models such as BERT~\cite{devlin2019bert}, often referred to as embedding models.
They allow condensing the contextual meaning of a text into a dense vector, termed embedding.
Using similarity metrics such as dot product, cosine similarity, or Euclidean distance allows for assessing the similarity of two input texts.
Embedding models are usually used for the similarity search in \gls{rag} systems~\cite{cuconasu2024power}, which we also do in our implementation.

\subsection{\glsxtrshort{llm} Agents}
\glspl{llm} have shown remarkable capabilities in solving complex tasks by decomposing them in a step-by-step fashion~\cite{wei2022chain} or by exploring multiple solution paths simultaneously~\cite{yao2024tree}.
Typically, these plans are generated iteratively by using the history of the previously generated steps to guide the generation of the next step.
Additionally, recent studies have shown the potential of providing \glspl{llm} access to external tools to boost their reasoning capabilities and add further knowledge.
This approach consists of prompting the \gls{llm} to interact with external tools to solve tasks, thus offloading computations from the \gls{llm} to specialized functions.
Notable examples of such tools include web browsers~\cite{nakano2021webgpt}, calculators~\cite{cobbe2021training}, and Python interpreters~\cite{gao2023pal}.
In practice, a tool is usually a Python function, which can be called during the interaction with the \gls{llm}.

The \gls{llm} agent paradigm~\cite{mialon2023augmented,openai2024function,yao2023react} combines the concepts of (i) external tool usage, (ii) the planning capabilities of \glspl{llm}, and adds a shared (iii) memory, to solve complex tasks.
Given an input task, an \gls{llm} agent uses its reasoning capabilities to decompose the task into a set of simpler subtasks.
For each subtask, the \gls{llm} finds and interacts with the set of tools to solve the subtask.
Then, based on the outcome of the current task and the history of previously executed subtasks, the \gls{llm} agent generates a new subtask and repeats the steps mentioned above or terminates if the original task is solved.
To instruct the processing, the outcome of the tool invocations and the history of the subtasks are stored in the memory, typically consisting in the \gls{llm} agent's own context.
Within this work, we apply the \gls{llm} agent paradigm to create the \theagent as an \gls{llm} agent for endpoint discovery.

A critical challenge for \gls{llm} agents is the accessibility to a set of common APIs and tasks for their evaluation, e.g., tested using benchmarks like API Bank~\cite{li2023apibank} or RestBench~\cite{song2023restgpt}.
API Bank is a benchmark to evaluate the tool use of an \gls{llm} consisting of a set of APIs exposed through a search engine.
Unfortunately, the available code of the benchmark is incomplete.
The RestBench benchmark contains a collection of tasks and endpoints expressed using the OpenAPI specification of Spotify and TMDB~\cite{song2023restgpt}.
As the currently most extensive available benchmark, we employ RestBench to validate our results.

OpenAPIs within \gls{llm} agents have been used in RestGPT~\cite{song2023restgpt} and Chain of Tools~\cite{shi2024chain}.
The former combines multiple \gls{llm} agents to solve complex tasks by interacting with a set of tools exposed using the OpenAPI specification. The latter solves an input query by framing the problem as a code generation task and interacts with the set of tools to generate Python code to solve the query.
In contrast, our \theagent does not directly interact with the endpoints stated in the OpenAPIs.
Instead, it filters and returns matching endpoints that can be used for subsequent processing.

Even when considering the similarity to the tool selection within \gls{llm} agents, the task of selecting a set of tools from a larger pool to solve a specific problem remains relatively underexplored~\cite{yuan2024craft}.
Existing research primarily focuses on the a priori selection of human-curated tools~\cite{parisi2022talm}, heuristic-based methods for tool selection~\cite{liang2024taskmatrix}, choosing the relevant tool by scoring each query against every tool using a similarity metric between user queries and API names~\cite{patil2023gorilla}, and embedding-based semantic retrieval using a combination of different vector databases~\cite{yuan2024craft}.
With our work, we contribute the analysis of preprocessing OpenAPIs into this corpus.

\section{Solution Design}
\label{sec:design}

We first introduce the general architecture to employ \gls{rag} for endpoint discovery.
As state-of-practice for service documentation, we then investigate how to chunk OpenAPIs as preprocessing for \gls{rag}.

\subsection{\glsfmtshort{rag} for Endpoint Discovery}

\gls{rag} comprises a preprocessing step ahead of the answer generation of an \gls{llm} to enrich the prompt with additional data.
Therefore, a retrieval component performs a semantic search based on some knowledge sources.
Usually, the semantic search is done by embedding similarity, and the data from the knowledge sources is reduced to small chunks to allow fine-grained information retrieval~\cite{lewis2020retrieval}.

\begin{figure}
    \centering
    \includegraphics[width=1.0\columnwidth]{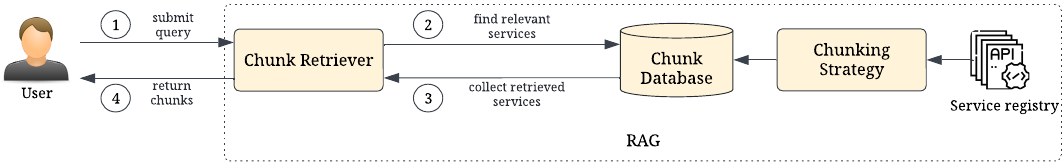}
    \caption{\glsxtrshort{rag} for Endpoint Discovery}
    \label{fig:soc-rag}
\end{figure}

\Cref{fig:soc-rag} depicts the application of \gls{rag} for endpoint discovery, i.e., the \therag.
Initially, the chunking strategy determines how the chunks are created from the OpenAPIs, i.e., how many chunks are created and what they contain.
Each chunk has an embedding as metadata for similarity search in addition to its content.
The chunking strategy specifies which data is used as input to the embedding model to create the embedding.
This input does not have to match the chunk content, e.g., it can be a summary instead of the entire content.
The chunks are finally stored in the chunk database.

For retrieval, the user submits in~\circled{1} a natural language query $q$ to the chunk retriever, which converts $q$ into the embedding $e$ using the same embedding model as for the chunk creation.
In~\circled{2}, the chunk retriever queries the chunk database using $e$.
The chunk database compares $e$ using a similarity metric with the embeddings of the service chunks contained in the database.
The results are the top $k$ most similar chunks according to the metric, which are then returned to the chunk retriever in~\circled{3}.
Finally, in~\circled{4}, the chunk retriever forwards the retrieved results to the user, who can add them to their prompt either manually or automatically through integration into their tooling.

The benefit of employing \gls{rag} is the insertion of only the gist of the available information, which allows picking more and only the most relevant information for the fix \gls{llm} context size.
A drawback is that, based on the retrieval algorithm, not all relevant information may be retrieved.
Further, fixing $k$ reveals the advantage of controlling the result size.
An alternative would be to return all chunks about a certain similarity threshold, introducing the question about the optimal cutoff.

\begin{figure}
    \centering
    \includegraphics[width=1.0\columnwidth]{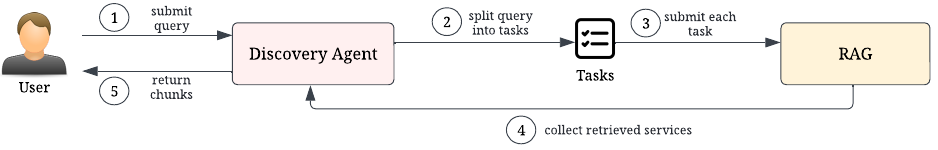}
    \caption{Overview of the \theagent Approach for Endpoint Discovery}
    \label{fig:soc-agent}
\end{figure}

\Cref{fig:soc-agent} shows how the \theagent extends on the \gls{rag} from \Cref{fig:soc-rag} shown in yellow hued.
Instead of passing $q$ to the \gls{rag}, the user submits it in~\circled{1} to the \theagent, which then iteratively decomposes $q$ into a set of fine-grained tasks in~\circled{2}.
Breaking down the query into smaller, more manageable tasks can potentially fill the gap between the coarse semantics of the query and the specificities in the services documentation.
In~\circled{3}, the \theagent submits each task to the \gls{rag} to retrieve the set of relevant chunks to solve the current task specifically.
Finally, in~\circled{4}, the \theagent collects the retrieval results of each individual task, filters them, and repeats~\circled{2} if $q$ needs further processing or returns the results to the user in~\circled{5}.

\subsection{OpenAPI Chunking Strategies}

A critical step in the \gls{rag} workflow is creating the chunks for the chunk database.
Embedding models typically have a limited input token size, and real-world service registries can contain tens of thousands of services, each containing multiple potentially lengthy endpoints due to detailed descriptions or extensive input and output schemas.
So, a single service might not fit into the context size of the embedding model or even exceed the limit of the \gls{llm} that further processes the output of the \gls{rag} system.
In addition, service documentation can also feature additional metadata that, while valuable for understanding service details, is not necessarily relevant for composing services to solve a query.

\begin{table}
    \caption{Implemented Chunking Strategies}
    \label{tab:approaches}
    \begin{tabulary}{\textwidth}{L|L|L|L}
        \makecell[c]{\textbf{Category}} & \makecell[c]{\textbf{Splitting}} & \makecell[c]{\textbf{Refinement}} & \makecell[c]{\textbf{Meta-Parameters}} \\
        \hline
        Token-based & No split & Token chunking & $m$ (model), $s$ (chunk size), $l$ (overlap) \\
                    & Endpoint split & Token chunking & $m$ (model), $s$ (chunk size), $l$ (overlap) \\
                    & Endpoint split & Remove examples & $m$ (model) \\
                    & Endpoint split & Relevant fields & $m$ (model) \\
                    & JSON split & Token chunking & $m$ (model), $s$ (chunk size), $l$ (overlap) \\
        \hline
        \glsxtrshort{llm}-based & Endpoint split & Query & $m$ (model) \\
                                & Endpoint split & Summary & $m$ (model)
    \end{tabulary}
    \centering
\end{table}

To determine advantageous chunking strategies, we employ the seven different chunking strategies presented in \Cref{tab:approaches}.
Input is always an OpenAPI specification, and output is a list of chunks.
The chunking strategies can be categorized into \textit{token-based} and \textit{\gls{llm}-based} strategies.
Each strategy consists of a \textit{splitting} method, which dissects the OpenAPI specification into a list of intermediate chunks, and another \textit{refinement} step, which converts the intermediate chunks to the final list of chunks.
In addition, there is the meta-parameter for the used embedding model~$m$.
For the token chunking refinement step, there is also the chunk size~$s$ in tokens and their overlap~$l$, i.e., how many tokens two consecutive chunks share, in tokens.

For the token-based approaches, we consider three main splitting methods.
The \textit{no split} method returns a single intermediate chunk for each OpenAPI containing the whole specification.
The \textit{endpoint split} divides the OpenAPI into one chunk per endpoint.
The \textit{JSON split} is a built-in LlamaIndex\footnote{\url{https://github.com/run-llama/llama_index}\label{foot:llamaindex}} splitting strategy tailored to files in the JSON format.
This strategy parses the JSON file and traverses it using depth-first search, collecting leaf nodes, i.e., key-value pair where the value is a primitive type, e.g., strings, numbers, etc..
During this traversal, the parser concatenates keys and values into single lines of text to create a comprehensive textual representation of each leaf node.

For the refinement, we implemented \textit{token chunking}, \textit{remove example}, and \textit{relevant field}.
The \textit{token chunking} splits each intermediate chunk into a list of fixed-size chunks of $s$ tokens respecting an overlap of $l$ tokens with the previous node.
The \textit{remove example} removes the \texttt{requestBody} and recursively all \texttt{examples} fields for each endpoint as these are typically lengthy but contribute little information.
The \textit{relevant field} extracts representative fields, i.e., service title, service description, endpoint verb, endpoint path, and endpoint description, which contribute much information but few tokens.

For the \gls{llm}-based processing strategies, we apply the endpoint split and a \textit{summary} (similar to~\cite{nogueira2019document}) and \textit{query} approach for refinement.
In the \textit{summary} approach, we prompt an \gls{llm} to generate a summary for each OpenAPI endpoint.
For the \textit{query} approach, we instruct the \gls{llm} to generate a possible query matching the OpenAPI endpoint, as this might be closer to a possible input query than the summary.
For both approaches, we only consider the \gls{llm} output for the embedding creation.
The chunk content remains the original OpenAPI endpoint information.
The no split and JSON split splitting methods can only be used with token chunking since all other refinement strategies rely on exactly one endpoint as an intermediate chunk.

\section{Evaluation}
\label{sec:evaluation}

To evaluate the \therag and the \theagent, we implement it as a fully operational prototype.
Then, we employ the RestBench~\cite{song2023restgpt} benchmark to validate it in a real-world setting.

\subsection{Implementation}
We implement the \therag and \theagent approaches as open-source prototypes based on the LlamaIndex library.\textsuperscript{\ref{foot:llamaindex}}
For the prototypes, we rely solely on OpenAPIs as the state-of-practice for service descriptions.
All sources and results are available online.\footnote{\url{https://doi.org/10.18419/darus-4605}\label{foot:sources}}

For the \therag, we focus on the components presented in \Cref{fig:soc-rag}.
At the first start, the system loads the OpenAPIs and applies a chunking strategy to create chunks and their embeddings for their later retrieval.
The chunks contain thereby the information from the OpenAPIs, e.g., a whole endpoint or a part of it.
A chunk embedding does not necessarily have to match the chunk's content; for example, the content can be the endpoint, and the embedding is created using a natural language summary of the endpoint.
Thus, the matching is performed based on the embedding, and the result returned is the chunk's content, which can include additional information not required for the matching process.
As the service database, we use FAISS, which allows the storage and the similarity search of chunks~\cite{douze2024faiss}.
We use a so-called QueryEngine from LlamaIndex for the chunk retriever, which allows us to query a chunk database based on textual input.

We realize the \theagent from \Cref{fig:soc-agent} using a LlamaIndex OpenAIAgent, which implements the \gls{llm} agent pattern for OpenAI's \glspl{llm}.
An OpenAIAgent takes a list of tools, i.e., Python functions with a name and a description as parameters, and interacts with these using the OpenAI API.
We implement two strategies for the tools.
In the first strategy (\textit{Query}), we use the \therag as input for a LlamaIndex QueryEngineTool, which allows the agent to interact with the \gls{rag} on demand.
This has the advantage of being a simple, straightforward implementation but may increase the token count as the results of the \gls{rag} are fed into the chat history, which is transferred to the \gls{llm} for the reasoning on this data.
The second strategy (\textit{Summary}) uses a \gls{rag} with chunks of the endpoint's verb, path, and summary as contents and for their embeddings.
We create the summary by instructing an \gls{llm} to create it based on the endpoint information, i.e., as in the summary chunking strategy.
This should reduce the token count, as the chunks are much smaller, as not all endpoint details are returned and processed.
To account for the same function as the first approach with the \therag and provide all information, we introduce a second tool, which takes the endpoint verb and path as input parameters and returns the whole endpoint information.
The complete data is only inserted into the history for indispensable endpoints.

To enable measuring the retrieved endpoints, we attach the endpoint information, i.e., verb and path, to each chunk as metadata.
For the endpoint split splitting strategies, we take the information from the endpoint.
For the other strategies, we first attach a list of all endpoints to the nodes before splitting and then filter on the endpoint paths in the final chunks after splitting.
So, for each chunk, we know to which endpoint or endpoints it relates to.

\subsection{Dataset and Metrics}

We evaluate our approach using the RestBench benchmark, covering the Spotify and TMDB OpenAPI specifications~\cite{song2023restgpt}.
With 40 endpoints for Spotify and 54 for TMDB, this benchmark is much more complex than usual \gls{soc} case studies containing usually just three to seven endpoints~\cite{pesl2024uncovering}.
Nevertheless, a holistic benchmark covering various domains is still missing (see \Cref{sec:related_work}).

RestBench contains 57 queries for Spotify and 100 for TMDB.
For each of these queries, a solution set of endpoints is given, i.e., one to four endpoints that must be called to fulfill the query.
For example, one query is \enquote{Who directed the top-1 rated movie?}
The solution contains the endpoints \enquote{GET /movie/top\_rated} and \enquote{GET /movie/\{movie\_id\}/credits.}

As embedding models, we employ OpenAI's \texttt{text-embedding-3-large}~\cite{openai2024embeddings} as one of the currently leading proprietary models.
As open-source models, we utilize BAAI/\theosmodelsmall~\cite{xiao2023bge}, which is relatively small while still producing reasonable results, allowing the model to be executed on commonly available hardware like laptops, and Nvidia's \theosmodellarge~\cite{lee2024nv} as one of the leading open-source models.
For the \gls{llm}, we use OpenAI's \texttt{gpt-4o-2024-05-13}.

We evaluate the quality of the retrieved information in terms of \textit{accuracy} and the token count of the returned result.
We measure the accuracy using standard information retrieval metrics, namely $\text{recall}=\frac{\text{TP}}{\text{TP}+\text{FN}}$, $\text{precision}=\frac{\text{TP}}{\text{TP}+\text{FP}}$, and $\text{F1}=\frac{2 \cdot \text{recall} \cdot \text{precision}}{\text{recall} + \text{precision}}$.

To correctly solve the query $q$, a service discovery approach's recall should ideally be maximal.
However, this could result in retrieving several irrelevant services, turning into a drop in precision.
The F1 score represents a balance between recall and precision.

\subsection{\glsxtrshort{rag}}

\begin{table}
    \caption{Results for the \therag for top $k = 10$ with the Spotify API. The overlap is in tokens. O represents OpenAI's \texttt{text-embedding-3-large}. S and N represent \theosmodelsmall and \theosmodellarge, respectively. TC is for token chunking, RE for remove examples, and RF for relevant fields. Recall, precision, and F1 are in percent. The highest values per column are marked in bold.}
    \label{tab:evaluation:rag:spotify}
    \begin{tabulary}{\textwidth}{L|L|L|R|R||*{3}{>{\raggedleft\arraybackslash}p{0.55cm}|}|*{3}{>{\raggedleft\arraybackslash}p{0.5cm}|}|*{2}{>{\raggedleft\arraybackslash}p{0.5cm}|}{>{\raggedleft\arraybackslash}p{0.5cm}}}
         &&&&& \multicolumn{3}{c||}{Recall} & \multicolumn{3}{c||}{Precision} & \multicolumn{3}{c}{F1} \\
        Category & Splitting & Refinement & \makecell[c]{$s$} & \makecell[c]{$l$} & \makecell[c]{O} & \makecell[c]{S} & \makecell[c]{N} & \makecell[c]{O} & \makecell[c]{S} & \makecell[c]{N} & \makecell[c]{O} & \makecell[c]{S} & \makecell[c]{N} \\
        \hline
        Character   & No       & TC & 1024 &  0  & 67 & 40 & 62 & 16 & 12 & \textbf{19} & 26 & 18 & 29 \\
                    & No       & TC & 1024 & 50  & 68 & 49 & 64 & 16 & 13 & \textbf{19} & 26 & 20 & 29 \\
                    & No       & TC & 8191 &  0  & 88 & 71 & 91 &  7 &  7 &  7 & 12 & 13 & 13 \\
                    & No       & TC & 8191 & 50  & 89 & 66 & 91 &  7 &  7 &  7 & 12 & 13 & 13 \\
                    & Endpoint & TC & 1024 &  0  & 70 & 75 & 76 & \textbf{19} & \textbf{19} & \textbf{20} & \textbf{29} & \textbf{31} & \textbf{31} \\
                    & Endpoint & TC & 1024 & 50  & 71 & 74 & 76 & \textbf{19} & \textbf{19} & \textbf{20} & \textbf{29} & \textbf{30} & \textbf{32} \\
                    & Endpoint & TC & 8191 &  0  & 73 & 75 & 76 & \textbf{19} & \textbf{19} & \textbf{20} & \textbf{29} & \textbf{30} & \textbf{31} \\
                    & Endpoint & TC & 8191 & 50  & 73 & 75 & 76 & \textbf{19} & \textbf{19} & \textbf{20} & \textbf{30} & \textbf{30} & \textbf{31} \\
                    & JSON     & TC & 1024 &  0  & 81 & 84 & 85 &  9 &  8 & 10 & 17 & 15 & 19 \\
                    & JSON     & TC & 1024 & 50  & 77 & 87 & 85 & 10 &  9 & 10 & 18 & 17 & 19 \\
                    & JSON     & TC & 8191 &  0  & \textbf{97} & \textbf{95} & \textbf{100} & 5 &  5 &  5 & 10 & 10 & 10 \\
                    & JSON     & TC & 8191 & 50  & \textbf{97} & \textbf{95} & \textbf{100} & 5 &  5 &  5 & 10 &  9 & 10 \\
                    & Endpoint & RE & N/A  & N/A & 72 & 75 & 73 & \textbf{18} & \textbf{19} & \textbf{19} & \textbf{29} & \textbf{30} & 30 \\
                    & Endpoint & RF & N/A  & N/A & 71 & 71 & 73 & \textbf{18} & \textbf{18} & \textbf{19} & \textbf{29} & 29 & 30 \\
        \hline
        \glsxtrshort{llm}   & Endpoint & Query   & N/A & N/A & 71 & 57 & 58 & \textbf{18} & 15 & 15 & 29 & 23 & 24 \\
                            & Endpoint & Summary & N/A & N/A & 72 & 74 & 67 & \textbf{18} & \textbf{19} & 17 & \textbf{29} & \textbf{30} & 27 \\
    \end{tabulary}
    \centering
\end{table}

\begin{table}
    \caption{Results for the \glsxtrshort{rag} for top $k = 10$ with the TMDB API. Schema as in \Cref{tab:evaluation:rag:spotify}.}
    \label{tab:evaluation:rag:tmdb}
    \begin{tabulary}{\textwidth}{L|L|L|R|R||*{3}{>{\raggedleft\arraybackslash}p{0.55cm}|}|*{3}{>{\raggedleft\arraybackslash}p{0.5cm}|}|*{2}{>{\raggedleft\arraybackslash}p{0.5cm}|}{>{\raggedleft\arraybackslash}p{0.5cm}}}
         &&&&& \multicolumn{3}{c||}{Recall} & \multicolumn{3}{c||}{Precision} & \multicolumn{3}{c}{F1} \\
        Category & Splitting & Refinement & \makecell[c]{$s$} & \makecell[c]{$l$} & \makecell[c]{O} & \makecell[c]{S} & \makecell[c]{N} & \makecell[c]{O} & \makecell[c]{S} & \makecell[c]{N} & \makecell[c]{O} & \makecell[c]{S} & \makecell[c]{N} \\
        \hline
        Character           & No       & TC      & 1024 &   0 &  1 & 14 &  7 & 33 & 17 & \textbf{38} &  2 & 15 & 11 \\
                            & No       & TC      & 1024 &  50 &  4 & 13 &  7 & \textbf{47} & \textbf{20} & \textbf{38} &  7 & 16 & 11 \\
                            & No       & TC      & 8191 &   0 & 30 & 17 & 15 & 19 &  5 &  9 & 23 &  8 & 11 \\
                            & No       & TC      & 8191 &  50 & 30 & 16 & 16 & 18 &  5 &  7 & 22 &  7 & 10 \\
                            & Endpoint & TC      & 1024 &   0 & 40 & 40 & 46 & 20 & 15 & 18 & 27 & 21 & \textbf{26} \\
                            & Endpoint & TC      & 1024 &  50 & 40 & 40 & 44 & 21 & 15 & 17 & 27 & 22 & \textbf{25} \\
                            & Endpoint & TC      & 8191 &   0 & 66 & 47 & 59 & 19 & 12 & 14 & \textbf{29} & 19 & 23 \\
                            & Endpoint & TC      & 8191 &  50 & 66 & 51 & 58 & 19 & 13 & 14 & \textbf{30} & 21 & 22 \\
                            & JSON     & TC      & 1024 &   0 & 44 & 44 & 46 & 18 & 12 & 16 & 26 & 19 & 24 \\
                            & JSON     & TC      & 1024 &  50 & 48 & 42 & 41 & 19 & 11 & 15 & 27 & 18 & 22 \\
                            & JSON     & TC      & 8191 &   0 & 61 & 60 & 50 &  8 &  8 &  6 & 14 & 14 & 11 \\
                            & JSON     & TC      & 8191 &  50 & 57 & 54 & 54 &  8 &  7 &  7 & 14 & 12 & 12 \\
                            & Endpoint & RE      &  N/A & N/A & \textbf{75} & 52 & \textbf{71} & 17 & 12 & 16 & 28 & 19 & \textbf{26} \\
                            & Endpoint & RF      &  N/A & N/A & 58 & 48 & 57 & 13 & 11 & 13 & 21 & 17 & 21 \\
        \hline
        \glsxtrshort{llm}   & Endpoint & Query   &  N/A & N/A & 56 & \textbf{65} & 46 & 13 & 15 & 10 & 20 & \textbf{24} & 17 \\
                            & Endpoint & Summary &  N/A & N/A & 69 & 59 & 65 & 16 & 13 & 15 & \textbf{29} & 22 & 24 \\
    \end{tabulary}
    \centering
\end{table}

\Cref{tab:evaluation:rag:spotify} shows the RestBench results for the \therag on the Spotify API.
In recall, the JSON split method performs exceptionally well, especially with a high chunk size $s$, as this approach densely packs the information from the JSON into the chunks by removing all formatting.
For precision and F1, the endpoint splitting approaches perform best because each chunk corresponds to precisely one endpoint.
Differences between the models are minor, except that the \theosmodelsmall performs worse for the no split approach.
We also tested $s = 2048$ and $s = 4096$, which are not reported here for space reasons.
We show $s = 1024$ because it is the default chunk size of LlamaIndex and $s = 8191$ because it is the maximum input token count for the OpenAI model.
It is worth mentioning that with an increasing chunk size, the token size of the returned result also increases.
Generally, a higher recall seems to correlate with a higher token count, e.g., no splitting with $s = 1024$, and $l=0$ has $4717$ tokens output on average.
In contrast, the JSON split has $10056$ with the same parameters, but this needs further analysis.
Due to length limitations, we cannot show the token count comparison and other values for top $k$ here.
We also tested top $k = 5$ and top $k = 20$.
Recall increases with a higher top $k$, but precision drops.
Additional data is available in the complementary material.\footref{foot:sources}

\Cref{tab:evaluation:rag:tmdb} presents the \therag RestBench results for the TMDB API.
The TMDB OpenAPI is more complex in length and extent than the Spotify OpenAPI.
In this case, the endpoint split-based approaches performs best in precision and F1.
The no split approaches achieve high values in precision due to their low value of true positives.

Overall, the endpoint split tends to outperform no splitting.
The JSON splitting benefits Spotify as the endpoints are already very dense, i.e., the endpoints do not contain examples, and schemas are only referenced.
Therefore, many endpoints can be condensed into one chunk.
This approach performs much worse for the lengthier endpoints in the TMDB API.
The summary refinement outperforms the query refinement, leading to the \theagent.

\subsection{\theagent}

\begin{table}[t]
    \caption{Results of the \theagent experiments. We set $\protect\text{top } k = 10$ and use OpenAI's \texttt{text-embedding-3-large} as the embedding model. Spotify and TMDB are the two test sets from the RestGPT benchmark. \glsxtrshort{rag} are the results for the summary chunking strategy from the \Cref{tab:evaluation:rag:spotify,tab:evaluation:rag:tmdb}. Query is the standard LlamaIndex QueryEngineTool to retrieve data from a \glsxtrshort{rag} system. The summary is our approach with a QueryEngineTool for summaries and a details-on-demand fetcher. Accuracy values are in percent. \#Token is the number of tokens per query averaged over all queries in the test set. The best value per row is marked in bold.}
    \label{tab:evaluation:agent}
    \begin{tabulary}{\textwidth}{*{7}{R|}R}
        \multicolumn{2}{c|}{} & \multicolumn{3}{c|}{Spotify} & \multicolumn{3}{c}{TMDB} \\
        \multicolumn{2}{c|}{} & \makecell[c]{\glsxtrshort{rag}} & \makecell[c]{Query} & \makecell[c]{Summary} & \makecell[c]{\glsxtrshort{rag}} & \makecell[c]{Query} & \makecell[c]{Summary} \\
        \hline
        Accuracy & Recall     & \textbf{71.92} & 63.70 &         66.44  & \textbf{69.33} & 43.11 &         46.67  \\
                 & Precision  &         18.42  & 67.39 & \textbf{70.29} &         15.60  & 45.97 & \textbf{50.97} \\
                 & F1         &         29.32  & 65.49 & \textbf{68.30} &         25.47  & 44.50 & \textbf{48.72} \\
        \hline
        \#Token & Prompt      &      4233.65  & 8606.87 & \textbf{3125.21} &     41001.46  & 65699.75 & \textbf{4544.57} \\
                & Completion  & \textbf{0.00} &  262.30 &          256.26  & \textbf{0.00} &   242.65 &          231.73  \\
                & Total       &      4233.65  & 8869.18 & \textbf{3411.47} &     41001.46  & 65942.40 & \textbf{4776.30} \\
    \end{tabulary}
    \centering
\end{table}

We present the RestBench results of the \theagent in \Cref{tab:evaluation:agent}.
For accuracy, we measure recall, precision, and F1 equally to the \therag experiments.
For the token count, we measure the actual tokens sent from the agent to the \gls{llm} from the agent as \textit{prompt}, the tokens received as \textit{completion}, and their sum as \textit{total}.
For the \gls{rag} approach, we accumulate the tokens of the retrieved chunks.

The results show that both agent approaches improve precision and F1 but reduce recall.
The Query approach increases the tokens in the prompt.
Contrarily, the Summary approach significantly outperforms the \gls{rag} and the query approach in the total token count.
The completion token count is by a magnitude smaller than the prompt token count for the agent approaches, which is relevant as completion tokens are usually more expensive than prompt tokens.
No \gls{llm} is invoked in the \gls{rag} approach, so the completion tokens are zero.

\subsection{Discussion}

We demonstrated the effectiveness of the \therag and the \theagent using our implementation.
They are able to retrieve large portions of relevant data while not revealing all relevant information in all cases.

To address RQ1, we implemented the \therag to apply \gls{rag} for endpoint discovery with seven chunking strategies and numerous parameter combinations.
We showed its effectiveness using the RestBench benchmark.
Overall, the ability to adequately reduce the token size to fit into the \gls{llm} context size while maintaining most of the relevant information is exhibited by the prototype.
Regarding the chunking strategies, endpoint split-based chunking strategies achieve favorable accuracies.
Limitations are primarily that the \gls{rag} results may not contain all relevant information, and the precision is low due to the retrieval of exactly $k$ chunks.
Additional research is needed to improve the retrieval performance further and prove the results in a generalized setting across multiple domains.

For RQ2, we introduced the \theagent, which transfers the \gls{llm} agent pattern to endpoint discovery.
Especially using Summary approach, the \theagent showed strong improvement over the \therag in terms of precision, F1, and token count.
Further research is needed to improve the decline in recall due to the processing through the \gls{llm}.

While we rely on the research benchmark RestBench for our results, which covers two extensive OpenAPIs, queries, and ground truth, it is still limited to these two services.
\therag systems in practice may operate on much larger datasets.
For the data processing, we rely on standard \gls{rag} implementations like LlamaIndex, which are already designed to operate on large amounts of data.
The performance evaluation, especially in larger real-world scenarios, remains open for future research.

The applicability of the \therag depends on the availability of service documentation.
We try to mitigate this issue by relying on widely adopted OpenAPI specifications, but this might not be valid for all domains.
A solution to consider is automatically generating service documentation using an \gls{llm}.

Another factor influencing the discovery is the quality of the OpenAPIs.
The discovery may fail if no descriptions, meaningful naming, or erroneous information is given.
This is not an issue of the approach, as a human developer would face the same problem, but it highlights the importance of high-quality documentation.

In addition to the presented chunking strategies, additional and more advanced strategies, e.g., CRAFT~\cite{yuan2024craft}, could be added to the \therag.
These could improve retrieval performance by combining multiple strategies or by creating a custom chunking strategy for a specific kind of service documentation.

Another advancement could also be creating a custom embedding model tailored explicitly to service descriptions and service description chunks.
This model may also be trained for one specific chunking strategy or intended use case.
Additionally, the \gls{rag} output may be trimmed to boost precision.
This could be done by, e.g., employing a similarity threshold.

The presented \theagent could be further improved to handle whole service compositions.
In this case, the agent would be extended by an additional component for the service composition, and the user would only submit their service composition task to the agent to retrieve the executable service composition solving this task.

Besides capabilities of the \gls{rag} system, resource consumption is a major issue in \gls{llm}-based systems.
The \therag only uses embedding models.
These are much more efficient than \glspl{llm}, resulting in costs in fractions of a cent per query.
In contrast, the \theagent requires significantly more resources, i.e., running RestBench in our experiments resulted in about \$50 of API fees.
Further work is needed to reduce this resource footprint.

\section{Concluding Remarks}
\label{sec:conclusion}

The service discovery challenge has been around for a long time in \gls{soc} to integrate different \glspl{is}.
With the application of automated \gls{llm}-based service composition approaches, the \gls{llm} input context limitations have become prominent, as the entire service documentation often does not fit into the input context, necessitating the preselection of relevant information.
To address this issue, we proposed an \therag, which facilitates semantic search based on state-of-the-practice OpenAPIs and reduces the input token size.
Further, we show an advanced integration through a \theagent, which can retrieve service details on demand to reduce the input token count further.
Our evaluation based on the RestBench benchmark shows that our approach is viable and performing.
Limitations are especially in the restriction of RestBench to two services of the entertainment domain.
We will address this in an extended version of this work.
Further improvements are in optimizing the implementation and extending the agent for additional tasks, e.g., whole service compositions.
We leave this for future work.

\begin{credits}
\subsubsection{\ackname}
This work was partially funded by the German Federal Ministry for Economic Affairs and Climate Action (BMWK) project Software-Defined Car (SofDCar) (19S21002).
The authors acknowledge support by the state of Baden-Württemberg through bwHPC.
This preprint has not undergone peer review (when applicable) or any post-submission improvements or corrections. The Version of Record of this contribution is published in Advanced Information Systems Engineering. CAiSE 2025. Lecture Notes in Computer Science, vol 15702. Springer, Cham., and is available online at \url{https://doi.org/10.1007/978-3-031-94571-7_8}.
\subsubsection{\discintname}
The authors Pesl and Aiello are listed as inventors of a patent~\cite{pesl2024verfahren}, which covers automated service composition using \glspl{llm} for the automotive domain.
\end{credits}

%
%
%
\bibliographystyle{splncs04}
\bibliography{bibliography}

\begin{thebibliography}{10}
\providecommand{\url}[1]{\texttt{#1}}
\providecommand{\urlprefix}{URL }
\providecommand{\doi}[1]{https://doi.org/#1}

\bibitem{achiam2023gpt}
Achiam, J., et~al.: {GPT}-4 technical report (2023),
  \url{https://arxiv.org/abs/2303.08774}

\bibitem{llama3modelcard}
AI@Meta: Llama 3 model card (2024),
  \url{https://github.com/meta-llama/llama3/blob/main/MODEL_CARD.md}

\bibitem{baresi2006distributed}
Baresi, L., Miraz, M.: A distributed approach for the federation of
  heterogeneous registries. In: ICSOC 2006. pp. 240--251. Springer (2006).
  \doi{10.1007/11948148_20}

\bibitem{bohn2008dynamic}
Bohn, H., Golatowski, F., Timmermann, D.: Dynamic device and service discovery
  extensions for {WS-BPEL}. In: ICSSSM 2008. pp.~1--6. IEEE (2008).
  \doi{10.1109/ICSSSM.2008.4598557}

\bibitem{cobbe2021training}
Cobbe, K., et~al.: Training verifiers to solve math word problems (2021),
  \url{https://arxiv.org/abs/2110.14168}

\bibitem{cuconasu2024power}
Cuconasu, F., et~al.: The power of noise: Redefining retrieval for {RAG}
  systems. In: SIGIR. vol.~47, pp. 719--729 (2024).
  \doi{10.1145/3626772.3657834}

\bibitem{curbera2002unraveling}
Curbera, F., et~al.: Unraveling the web services web: an introduction to
  {SOAP}, {WSDL}, and {UDDI}. IEEE Internet Computing  \textbf{6}(2),  86--93
  (2002). \doi{10.1109/4236.991449}

\bibitem{devlin2019bert}
Devlin, J., Chang, M.W., Lee, K., Toutanova, K.: {BERT}: Pre-training of deep
  bidirectional transformers for language understanding. In: NAACL-HLT 2019.
  pp. 4171--4186 (2019)

\bibitem{douze2024faiss}
Douze, M., et~al.: The {Faiss} library (2024),
  \url{https://arxiv.org/abs/2401.08281}

\bibitem{fan2023large}
Fan, A., et~al.: Large language models for software engineering: Survey and
  open problems (2023), \url{https://arxiv.org/abs/2310.03533}

\bibitem{10.1007/978-3-540-24593-3_5}
Fikouras, I., Freiter, E.: Service discovery and orchestration for distributed
  service repositories. In: ICSOC 2003. pp. 59--74. Springer (2003).
  \doi{10.1007/978-3-540-24593-3_5}

\bibitem{gao2023pal}
Gao, L., et~al.: Pal: Program-aided language models. In: International
  Conference on Machine Learning. pp. 10764--10799. PMLR (2023)

\bibitem{kim2024leveraging}
Kim, M., Stennett, T., Shah, D., Sinha, S., Orso, A.: Leveraging large language
  models to improve {REST} {API} testing. In: ICSE. vol.~44, pp. 37--41 (2024).
  \doi{10.1145/3639476.3639769}

\bibitem{lee2024nv}
Lee, C., et~al.: Nv-embed: Improved techniques for training llms as generalist
  embedding models (2024), \url{https://arxiv.org/abs/2405.17428}

\bibitem{lemos2015web}
Lemos, A.L., Daniel, F., Benatallah, B.: Web service composition: A survey of
  techniques and tools. ACM Comput. Surv.  \textbf{48}(3) (dec 2015).
  \doi{10.1145/2831270}

\bibitem{lewis2020retrieval}
Lewis, P., et~al.: Retrieval-augmented generation for knowledge-intensive {NLP}
  tasks. In: NeurIPS. vol.~33, pp. 9459--9474. Curran Associates (2020)

\bibitem{li2023apibank}
Li, M., et~al.: {API-Bank}: A comprehensive benchmark for tool-augmented
  {LLMs}. In: EMNLP. Association for Computational Linguistics (2023).
  \doi{10.18653/v1/2023.emnlp-main.187}

\bibitem{liang2024taskmatrix}
Liang, Y., et~al.: {Taskmatrix.AI}: Completing tasks by connecting foundation
  models with millions of {APIs}. Intelligent Computing  \textbf{3}, ~0063
  (2024). \doi{10.34133/icomputing.0063}

\bibitem{mialon2023augmented}
Mialon, G., et~al.: Augmented language models: a survey (2023),
  \url{https://arxiv.org/abs/2302.07842}

\bibitem{nakano2021webgpt}
Nakano, R., et~al.: {WebGPT}: Browser-assisted question-answering with human
  feedback (2021), \url{https://arxiv.org/abs/2112.09332}

\bibitem{nogueira2019document}
Nogueira, R., Yang, W., Lin, J., Cho, K.: Document expansion by query
  prediction (2019), \url{https://arxiv.org/abs/1904.08375}

\bibitem{openai2024function}
OpenAI: Function calling and other {API} updates (Jun 2024),
  \url{https://openai.com/index/function-calling-and-other-api-updates/}, last
  accessed 2024-07-18

\bibitem{openai2024contextsize}
OpenAI: {GPT-4 Turbo in the OpenAI API}.
  \url{https://help.openai.com/en/articles/8555510-gpt-4-turbo-in-the-openai-api}
  (2024), last accessed 2024-11-19

\bibitem{openai2024embeddings}
OpenAI: New embedding models and {API} updates (Jan 2024),
  \url{https://openai.com/blog/new-embedding-models-and-api-updates}, last
  accessed 2024-07-18

\bibitem{parisi2022talm}
Parisi, A., Zhao, Y., Fiedel, N.: Talm: Tool augmented language models (2022),
  \url{https://arxiv.org/abs/2205.12255}

\bibitem{patil2023gorilla}
Patil, S.G., Zhang, T., Wang, X., Gonzalez, J.E.: Gorilla: Large language model
  connected with massive {APIs} (2023), \url{https://arxiv.org/abs/2305.15334}

\bibitem{pesl2024verfahren}
Pesl, R.D., Klein, K., Aiello, M.: {Verfahren} zur {Nutzung} von unbekannten
  neuen {Systemdiensten} in einer {Fahrzeuganwendung} (2024), {Patent
  DE102024108126A1}

\bibitem{pesl2024uncovering}
Pesl, R.D., Stötzner, M., Georgievski, I., Aiello, M.: Uncovering {LLMs} for
  service-composition: Challenges and opportunities. In: ICSOC 2023 WS.
  Springer (2024). \doi{10.1007/978-981-97-0989-2_4}

\bibitem{pesl2024compositio}
Pesl, R.D., et~al.: {Compositio} {Prompto}: An architecture to employ large
  language models in automated service computing. In: ICSOC 2024. Springer
  (2024)

\bibitem{radford2019better}
Radford, A., Wu, J., Amodei, D., Amodei, D., Clark, J., Brundage, M.,
  Sutskever, I.: Better language models and their implications. OpenAI blog
  \textbf{1}(2) (2019), \url{https://openai.com/index/better-language-models/},
  last accessed 2024-11-28

\bibitem{radford2018improving}
Radford, A., et~al.: Improving language understanding by generative
  pre-training (2018)

\bibitem{santana2006upnp}
Santana, J.M.S., Petrova, M., Mahonen, P.: {UPnP} service discovery for
  heterogeneous networks. In: IEEE PIMRC. vol.~17, pp.~1--5. IEEE (2006)

\bibitem{shi2024chain}
Shi, Z., et~al.: Chain of tools: Large language model is an automatic
  multi-tool learner (2024), \url{https://arxiv.org/abs/2405.16533}

\bibitem{10.1007/978-3-031-57853-3_3}
Soki, A.T., Siqueira, F.: Discovery of {RESTful} {Web} services based on the
  {OpenAPI} 3.0 standard with semantic annotations. In: AINA. pp. 22--34.
  Springer (2024). \doi{10.1007/978-3-031-57853-3_3}

\bibitem{song2023restgpt}
Song, Y., et~al.: {RestGPT}: Connecting large language models with real-world
  applications via restful {APIs} (2023),
  \url{https://arxiv.org/abs/2306.06624}

\bibitem{thones2015microservices}
Th{\"o}nes, J.: Microservices. IEEE software  \textbf{32}(1),  116--116 (2015).
  \doi{10.1109/MS.2015.11}

\bibitem{vaswani2017attention}
Vaswani, A., et~al.: Attention is all you need. NeurIPS  \textbf{30} (2017)

\bibitem{wei2022chain}
Wei, J., et~al.: Chain-of-thought prompting elicits reasoning in large language
  models. NeurIPS  \textbf{35},  24824--24837 (2022)

\bibitem{xiao2023bge}
Xiao, S., Liu, Z., Zhang, P., Muennighoff, N.: C-pack: Packaged resources to
  advance general chinese embedding (2023),
  \url{https://arxiv.org/abs/2309.07597}

\bibitem{yao2023react}
Yao, S., et~al.: React: Synergizing reasoning and acting in language models
  (2023), \url{https://arxiv.org/abs/2210.03629}

\bibitem{yao2024tree}
Yao, S., et~al.: Tree of thoughts: Deliberate problem solving with large
  language models. NeurIPS  \textbf{36} (2024)

\bibitem{yuan2024craft}
Yuan, L., et~al.: {CRAFT}: Customizing {LLMs} by creating and retrieving from
  specialized toolsets (2024), \url{https://arxiv.org/abs/2309.17428}

\end{thebibliography}
\end{document}